# Nickel-cobalt-titanate thin films – new sustainable magnetic oxides

Yukari Fujioka[1], Johannes Frantti[*,1], Christopher Rouleau[2], Alexander Puretzky[2], and Harry M. Meyer[3].

[1] Finnish Research and Engineering, Jaalaranta 9 B 42, Helsinki 00180, Finland

[2] Center for Nanophase Materials Sciences, Oak Ridge National Laboratory, Oak Ridge, Tennessee, 37831, USA

[3] Oak Ridge National Laboratory, Oak Ridge, Tennessee, 37831, USA

[*] Corresponding author. Email: Johannes.frantti@fre.fi

**Abstract**

Single phase nickel-cobalt-titanate thin films with a formula $A_{1+2x}Ti_{1-x}O_3$, where $A$ is $Ni^{2+}$, $Co^{2+}$ and -0.25<$x$<1, were grown by pulsed laser deposition on sapphire substrates. There is a large window in which both Ni/Co ratio and $x$ can be chosen independently. In the prototype ilmenite and corundum structures one third of the octahedra are vacant. The reported structure is obtained by filling vacant ($x$>0) or emptying filled ($x$<0) octahedra. When $x$ = 1 all octahedra are filled. Two factors controlling the magnetism and crystal distortion are identified. First is a direct overlap between the adjacent cation *d*-orbitals resulting in a bond formation and magnetic interactions between the cations. This is most clearly revealed as a crystal distortion in the $x \approx 0$ compositions with approximately equal amounts of Ni and Co: the distortion of the $x \approx 0$ compound is a function of Ni/Co ratio. The second factor is $x$, which controls the cation shift towards a vacant octahedron. The displacement decreases and the symmetry increases with decreasing Ti content as was revealed by x-ray diffraction and Raman spectroscopy. When all octahedra are filled the cations prefer octahedron center positions. Also the number density of cations has increased by a factor of 50% when compared to the ilmenite structure. The number density ratios of Ni/Co cations between $x$=1 and $x$=0 compounds is 3. The Raman and x-ray diffraction data collected on samples with $x$ = 1 or close to 1 are interpreted in terms of $P6_3/mmc$ space group.

1. Introduction

One route to look for alternative materials is to systematically modify the composition of the prevailing materials. Correspondingly, the number of work dedicated to perovskites is immense. Another, less studied alternative materials are ilmenite oxides (1-8), which are structurally similar to the lithium niobate (9) and the corundum structures (2). There are two structural features which makes the ilmenite and corundundum structures particularly appealing, presence of short cation-cation distances and vacant octahedra. Nickel-cobalt-titanate $(Ni_{1-x}Co_x)TiO_3$ (NCT) ilmenite solid-solution exhibits triclinic structure and spontaneous magnetization and so possesses a potential for electronic applications requiring ferromagnetic properties (6). The shortest cation-cation distances in NCT are 2 Å, to be compared to 3.9 Å found in the perovskite lead titanate. Vacant sites open up new tailoring possibilities (examples are given in ref. 10), which is exploited in the present study. Thin film properties often deviate from the corresponding bulk properties. Especially the layer composition can be controlled which makes structures not stable in a bulk form possible. Lattice mismatch between the film and substrate causes strain, which is one method to develop ferroelectric and ferromagnetic NCT thin films.

Common reasons to develop sustainable materials are to replace toxic materials, find economical alternatives or to find technologically better materials. The restriction of the use of Hazardous Substances (RoHs) requires heavy metals such as lead, mercury, cadmium, and hexavalent chromium to be substituted by safer alternatives (11). Heavy metals are raw material for many high performance functional compounds applied in electronics. Rare earth elements (REE) are widely used in strong permanent magnets. Triggered by the REE price peak in 2011 efforts were launched to find sustainable and cheaper alternative materials (12). Surface oxidation is a serious technological problem in many REE magnets and evidently even more critical issue for thin films (13). Though REE magnets have excellent magnetic properties, there are numerous applications which do not require extreme magnetization values. This opens up routes to develop new sustainable materials.

In this paper we report the structural properties of NCT and NCT-derived thin films. First Ti-rich films are addressed after which the (Ni,Co)-rich films are discussed. A structural model summarizing the properties of both types of films is given.

2. Why ilmenites – structural aspects

Ilmenite $ABO_3$ oxides possess unique structural features enabling rich functionality. The structure, Fig. 1, is layered; $AO_6$ and $BO_6$ octahedra alternate along the hexagonal $c$ axis (14). Each layer consists of two slightly separated, two-dimensional triangular cation nets. The separation is due to the electrostatic repulsion which is minimized by displacing the cations towards vacant octahedra, see Fig. 1 (a). As is discussed below, the vacant octahedra have a crucial role for stabilizing a new structure. The octahedra share edges in the layers, see Fig. 1 (b), between the layers the octahedra are connected by sharing faces, Fig. 1 (a). Besides oxygen and two-oxygen mediated magnetic interactions (15) the direct overlap between the adjacent $d$-orbitals in ilmenites results in a bond formation and additional magnetic interactions between the cations. This is in contrast to the perovskite structure in which octahedra share only corners. The short cation-cation distances are actually a factor behind the Pauling's third rule according to which the existence of edges and particularly of faces, common to two anion polyhedra in a coordinated structure decreases its stability (16). This implies that the structures possessing these properties are generally not energetically favorable.

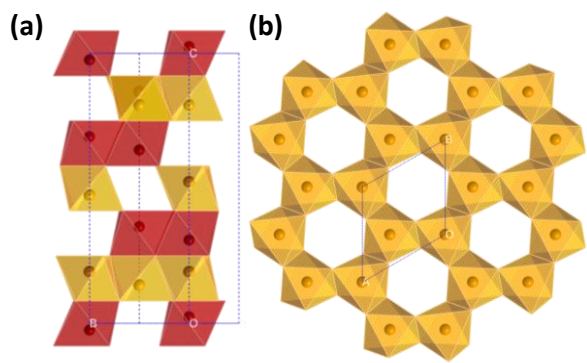

Fig. 1. The ilmenite structure. Red and gold-colored octahedra are Ti- and Ni/Co octahedra, respectively. Panel (a) gives an *bc*-plane projection and panel (b) shows an *ab*-plane projection. For clarity, only Ni/Co-octhaedron network is shown in panel (b). The structure is consisted of octahedra of which 1/3 are vacant.

Neutron and synchrotron x-ray diffraction (6) and Raman scattering (17) studies revealed that NCT in a bulk form possesses a lower crystal symmetry than $R\bar{3}$ and also two magnetic phases absent in NiTiO$_3$ and CoTiO$_3$. NiTiO$_3$ (18) and CoTiO$_3$ (19) are reported to have $R\bar{3}$ symmetry. $x$ = 0.50 NCT is ferromagnetic below 70 K (6,7). Both crystal distortion and magnetism were assigned to the interactions between cations, there being interactions through octahedral edges via $t_{2g}$ orbital overlap between Ni$^{2+}$-Ni$^{2+}$, Co$^{2+}$-Co$^{2+}$, and Ni$^{2+}$-Co$^{2+}$ cation pairs within the *ab*-plane. There is also an orbital overlap interaction between the layers. Each Ni/Co-layer octahedra share a face with Ti-layer octahedra only on one side of the layer, see Fig. 1(a). An implication of this is that the half-filled $t_{2g}$ Co orbital can ferromagnetically couple with an empty Ti $t_{2g}$ orbital if there is a partially occupied state at Ti (e.g., some Ti-cations are in a Ti$^{3+}$ state) which is orthogonal to the empty Ti state (20). Interestingly, a reversible structural transformation from ilmenite structure with alternating Ti-Ni/Co layers to the corundum structure with mixed Ti/Ni/Co layers took place during heating in $x$ = 0.50 sample, indicating that the interactions between Ni/Co and Ti cations are strongly temperature dependent (17). Thus, by adjusting the Ni-Co-Ti distribution one can change the magnetic properties and crystal distortion or even crystal structure. For this purpose pulsed laser deposition (PLD) is an excellent method. We note that similar features to the NCT are also seen in iron oxides, including a solid solution of $\alpha$-Fe$_2$O$_3$ and FeTiO$_3$ (21).

### 3. Experimental

NCT thin films were deposited by PLD technique at the Center for Nanomaterials Sciences (CNMS) on sapphire substrates with 0001 (MTI Corporation) and 11$\bar{2}$0 and 10$\bar{1}$0 (Mateck) planes. The substrate planes are labeled using the hexagonal indices (*hkil*), where the redundant index *i* = -*h* -*k*. Reflections from thin films are labeled using the three index notation. Three targets, NiTiO$_3$, CoTiO$_3$ and NiCoO$_4$ were ablated repeatedly. Shot ratios and total number of shots were adjusted to control the composition and film thickness and the targets were rotated during ablation. Substrates were in-situ heated. Laser beam wavelength was 248 nm and the pulse repetition rate was 10 Hz. Table 1 summarizes the deposition conditions and thin film composition as detected by X-ray photoelectron spectroscopy (XPS).

Table 1. Summary of the deposition conditions and composition as determined through XPS measurements. $T$ is substrate temperature, $pO_2$ is $O_2$ pressure and Fluence is the laser beam pulse energy per area. The composition of the sample H was estimated from the known compositions of three thin films deposited under similar conditions.

| Sample | Composition | $pO_2$ (mTorr) | $T$ (°C) | Fluence (Jcm$^{-2}$) | Substrate |
|---|---|---|---|---|---|
| A | $Ni_{0.32}Co_{0.24}Ti_{1.35}O_3$ | 5 | 475 | 4.8 | $Al_2O_3$ (11$\bar{2}$0) |
| B | $Ni_{0.46}Co_{0.18}Ti_{1.35}O_3$ | 5 | 475 | 3.0 | $Al_2O_3$ (11$\bar{2}$0) |
| C | $Ni_{0.46}Co_{0.81}Ti_{0.82}O_3$ | 10 | 590 | 3.0 | $Al_2O_3$ (0001) |
| D | $Ni_{0.49}Co_{0.92}Ti_{0.70}O_3$ | 10 | 500 | 3.0 | $Al_2O_3$ (0001) |
| E | $Ni_{1.07}Co_{1.72}O_3$ | 10 | 610 | 3.0 | $Al_2O_3$ (0001) |
| F | $Ni_{1.07}Co_{1.72}O_3$ | 10 | 610 | 3.0 | $Al_2O_3$ (10$\bar{1}$0) |
| G | $Ni_{1.07}Co_{1.72}O_3$ | 10 | 610 | 3.0 | $Al_2O_3$ (11$\bar{2}$0) |
| H | $Ni_{0.78}Co_{1.02}Ti_{0.46}O_3$ | 10 | 610 | 3.0 | $Al_2O_3$ (0001) |
| I | $Ni_{0.78}Co_{1.02}Ti_{0.46}O_3$ | 10 | 610 | 3.0 | $Al_2O_3$ (10$\bar{1}$0) |
| J | $Ni_{0.57}Co_{0.69}Ti_{1.17}O_3$ | 10 | 610 | 3.0 | $Al_2O_3$ (0001) |

Raman measurements were performed using a Jobin-Yvon T64000 spectrometer consisting of a double monochromator coupled to a third monochromator stage with 1800 grooves per millimeter grating (double substractive mode). A liquid nitrogen cooled charge-coupled device detector was used to count photons. All measurements were carried out under a microscope in backscattering configuration. Raman spectra were excited using a continuous wave solid-state laser (wavelength 532 nm).

X-Ray Diffraction (XRD) data on thin films were collected by Panalytical X'Pert Pro MPD diffractometer with a CuK$\alpha$ source, Ni filter and X'Celerator detector.

XPS measurements were conducted for the determination of the film composition as a function of depth (XPS, K-Alpha XPS system, Thermo Fisher Scientific) equipped with a monochromated Al-K$\alpha$ source ($h\nu$ = 1486.6 eV) and argon ion sputtering.

## 4. Results
### 4.1. Ti-rich and slightly (Ni,Co)-rich films

Figure 2 shows XRD patterns measured from NCT thin films. Except for sample B, only (00*l*) type peaks were detected, indicating a preferred orientation of the films grown on (0001) sapphire. In the case of prototype ilmenite structure this would imply that the hexagonal *c* axis is perpendicular to the substrate plane. Since strain and non-prototype composition can lower the symmetry this may no longer be true. Below we discuss the results in terms of hexagonal axes, keeping in mind crystal distortions lowering the symmetry. Samples A and B were grown on (11$\bar{2}$0) sapphire substrates, the most apparent difference being the larger laser pulse energy applied in the growth of film A which could be the reason why, somewhat surprisingly, the film had preferred 00*l* orientation. Next section gives results for other films grown with similar pulse energies as samples B, C and D, which also indicate that films grown with modest pulse energies on (0001) sapphire substrates have preferred 00*l* orientation. It is also shown that except for the sample B, films grown on (11$\bar{2}$0) sapphire substrates were 00*l* oriented.

Though the peak positions match with the ilmenite structure the intensity ratios of the peaks do not correspond to a case in which (Ni,Co) and Ti are segregated in their own layers (Model 1), as is shown by Table 2. Since the data for samples A, C and D allow only the determination of the *c*-axis length, we assumed that the space group symmetry is $R\bar{3}$ and constrained the *a* axis length of NiTiO$_3$, 5.031 Å (6). Also the fractional coordinates of ions were constrained. Two simple models were tested, Model 1 assumes that (Ni$^{2+}$,Co$^{2+}$) and Ti$^{4+}$ ions are segregated at their own layers as in the ideal ilmenite structure. Model 2 assumes that all layers are compositionally equivalent and contain cations in the composition ratios given in

Table 1. Models 1 and 2 are given in Table 4 in Appendix 1. Occupancies for samples A, C and D are given in Table 4. The *c* axis length was 13.8114, 14.5800 and 14.5846 Å for samples A, C and D, respectively.

Table 2 gives intensity ratios when all cations are assumed to be mixed in the same layer in their statistical ratios (Model 2).

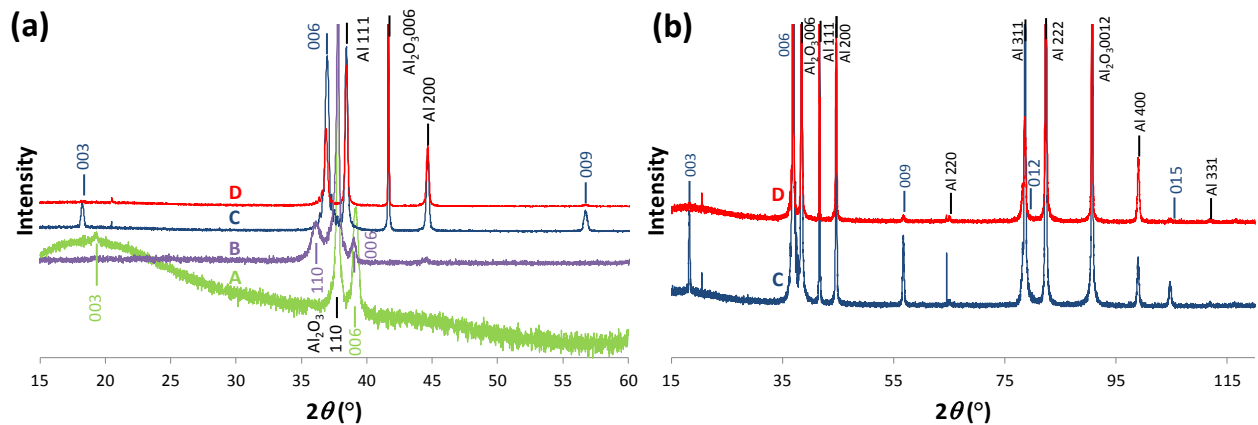

Fig. 2. (a) XRD patterns measured from thin films A, B, C and D. The reflections from all films were indexed by a single ilmenite structure. Also reflections from the sapphire substrates and the aluminum sample stage are given. To confirm the single phase nature and to observe (0012) and (0015) reflections a larger scan was carried out for the samples C and D.

Table 2. Measured and modeled intensity ratios of the 00*l* reflections for samples A, C and D.

|   | Sample A | | | Sample C | | | Sample D | | |
|---|---|---|---|---|---|---|---|---|---|
| *l* | Measured | Model 1 | Model 2 | Measured | Model 1 | Model 2 | Measured | Model 1 | Model 2 |
| 3 | 0.12(3) | 1 | 0 | 0.03(1) | 1 | 0 | 0.017(2) | 0.999 | 0 |
| 6 | 1 | 0.439 | 1 | 1 | 0.624 | 1 | 1 | 1 | 1 |
| 9 | - | 0.000 | 0 | 0.03(1) | 0.000 | 0 | 0.021(1) | 0 | 0 |
| 12 | - | 0.013 | 0.040 | 0.26(6) | 0.000 | 0.012 | 0.41(2) | 0.000 | 0.011 |
| 15 |  | 0.001 | 0 | 0.013(3) | 0.001 | 0 | 0.02(2) | 0.001 | 0 |

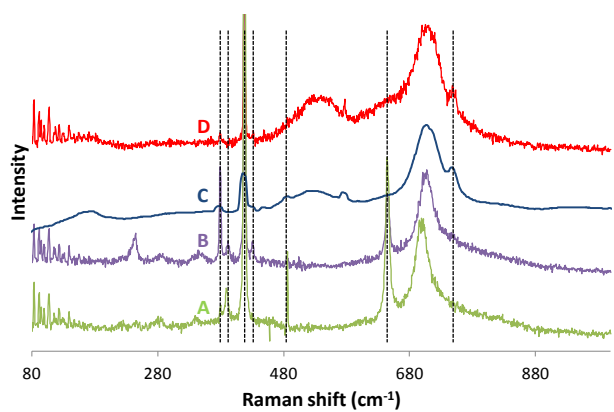

Fig. 3. Raman spectra measured from films A, B, C and D. Spectra measured from samples A and B are reminiscent to the ilmenite structure, whereas spectrum measured from sample D lacks the weak details between 200 and 400 cm$^{-1}$. Dotted lines indicate sapphire lines from substrate.

Though the non-zero intensities of the 003*n*, *n* odd, peaks show that Model 2 cannot be precisely correct, it gives some features correctly in the case of sample D: the intensities of the 006 and 0012 reflections are strong, whereas the 003, 009 and 0015 reflections are all weak even though not exactly zero. The tendency of 003 peak to get weaker with increasing (Ni+Co)/Ti ration is seen. Though Model 1 explains the presence of the 003 peak, it should be modified since the predicted 003 intensity is larger than observed.

Raman scattering measurements is in line with the XRD results. In the prototype ilmenite and corundum structures the Raman active Brillouin zone centre modes transform as $5A_g+5E_g$ and $2A_{1g}+5E_g$, respectively. Ten Raman peaks are expected for the ilmenite and seven for the corundum phase. By consulting ref. 22 it is seen that if a single crystal is oriented so that both the polarization vector of the laser light and the polarization direction of the analyzed component of the scattered are parallel to the hexagonal $c$ axis only $A_g$ ($A_{1g}$) modes are detected. If both the incoming and scattered light polarization directions are perpendicular to each other and to the $c$ axis, only $E_g$ modes are allowed. In the case of the $c$ axis oriented films both $A_g$ ($A_{1g}$) and $E_g$ symmetry modes are observed. In NiTiO$_3$ thin films, possessing the ilmenite structure, the five modes at 193.1, 247.6, 395.2, 485.2 and 706.8 cm$^{-1}$ were assigned to $A_g$ symmetry and the five modes at 230.5, 292.0, 345.5, 465.3 and 609.6 cm$^{-1}$ were assigned to $E_g$ symmetry (23). Fig. 3 shows that the samples possessing large 003 peak also have several Raman peaks between 200 and 400 cm$^{-1}$, in addition to the strong peak observed at around 700cm$^{-1}$. The spectra collected from samples A and B are similar to the spectra collected on bulk NCT samples (17) and are characteristic to the ilmenite structure. Spectrum measured from sample D is significantly simpler and is characterized by a strong band between 450 and 800 cm$^{-1}$, which is also seen in sample C. Raman scattering results are consistent with the XRD results shown in Fig. 2 and are interpreted in terms of a cation-mixing structural transformation.

XPS measurements as a function of film thickness indicated that the film composition was essentially constant, as shown in Fig. 4. The valence state of Ni, Co and Ti cations were found to be 2+, 2+ and 4+, respectively. Despite there being a large charge difference between the cations present in the same layer, no evidence for the lithium niobate or other charge-ordered structure was found. Large charge difference is often accompanied by cation ordering. Nearly random distribution of cations is probably related to relative low *in-situ* substrate temperatures, see Table 1. Higher substrate temperatures would increase cation diffusion during film growth.

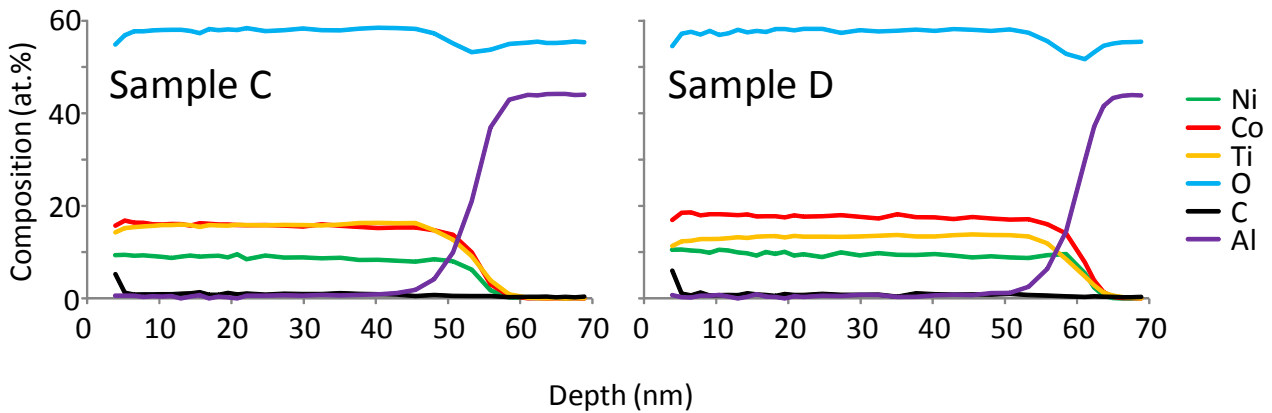

Fig. 4. The composition of the films C and D as a function of thickness. The composition is seen to be homogeneous throughout the film thickness. Substrate is marked by an increase in Al-signal.

In conclusion, the titanium-rich films, samples A and B, possess the ilmenite-derived structure even though the composition does not follow the (Ni,Co)TiO$_3$ stoichiometry. Samples C and D show that the distortion related to the ilmenite structure diminished with decreasing Ti content. Though both Models 1 and 2 qualitatively describe many features, they should be modified. We return to this in section 4.3.

## 4.2. (Ni,Co)-rich films

To gain better understanding of the structural changes a set of Ti-deficient thin film were grown. Fig. 5 shows XRD patterns and Raman spectra measured from sample E, F, G, H, I and J with compositions tabulated in Table 1, the difference between samples H and I being different film orientation due to the different substrate orientation.

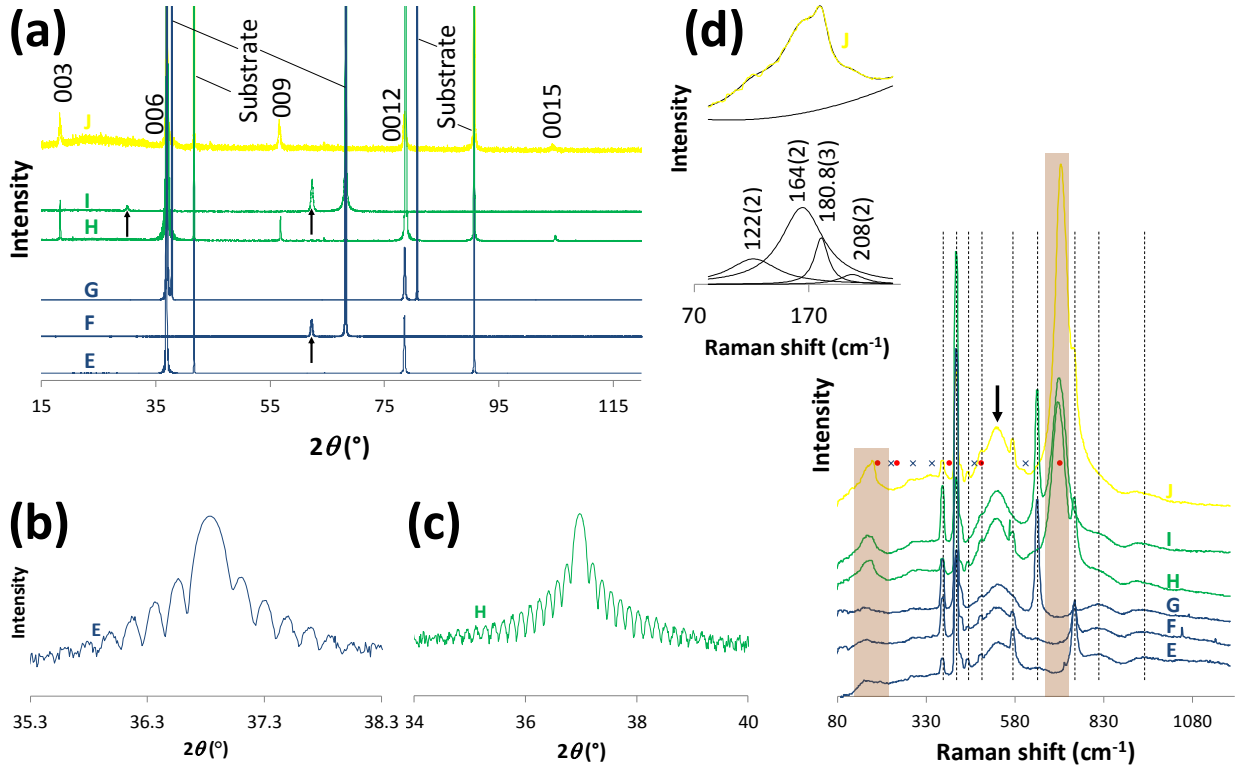

Fig. 5. (a) XRD pattern measured from samples E, F, G, H, I and J. Arrows indicate the reflections observed at 30° and 62° in the sample I. The arrow at around 62° indicates the only reflection seen from sample F. Panels (b) and (c) show the XRD intensity of the 006 reflection of the E and F samples, respectively. Subsidiary minima are clearly revealed, consistently with the idea that the reflections originate from a single phase. The film thicknesses were estimated to be 44.3 (sample E) and 53.7 nm (sample H). Appendix 2 gives the details of thickness estimation. Panel (d) gives the corresponding Raman spectra. Red spheres and blue crosses give the positions of the $A_{1g}$ and $E_g$ symmetry modes reported in NiTiO$_3$ thin films, ref. 23. Peaks from sapphire substrate are indicated by dotted vertical lines. The arrow indicates the mode remaining at around 530 cm$^{-1}$ when Ti-content diminishes to zero. The rectangular light purple shadows indicate the spectral regions undergoing the most apparent changes as a function of Ti/(Ni+Co) ratio. The inset in panel (d) gives the peak fit of the low-frequency part of the sample J.

Only 00$l$ reflections are observed from samples E, H and J deposited on 0001-oriented sapphire substrates. The XRD pattern Fig. 5(a), and Raman spectra, Fig. 5(d), of the sample J are consistent with the Ni$_{1-x}$Co$_x$TiO$_3$ spectrum measured from a ceramic bulk sample with $x$ = 0.50. The peak at around 193.1 cm$^{-1}$ was also assigned to $A_{1g}$ symmetry (23). The film J exhibits similar low-frequency distortion as was found in bulk Ni$_{1-x}$Co$_x$TiO$_3$ ceramics (17), as shown in the inset of Fig. 5(d). The bulk sample with $x$ = 0.50 had a low frequency peaks at 123, 153, 168 and 184 cm$^{-1}$. Since the scattering volume of thin films is very small, we used a single monochromator set-up to have a better signal-to-noise ratio, which compromises with the resolution. Thus, the peak at 164 cm$^{-1}$ might be a doublet corresponding to the peaks found at 153 and 168 cm$^{-1}$ in bulk sample. The low-frequency band suggests that the symmetry of the film is lower than $R\bar{3}$, in the line discussed in ref. 17. The distortion is characteristic to Ni$_{1-x}$Co$_x$TiO$_3$ solid-solutions: Ni and Co rich compounds do not have this 3$d$-orbital overlap related band.

The XRD pattern of the sample E exhibits only 006 and 0012 reflections, Fig. 6(a). Interestingly, also sample G is 00$l$ oriented though it is grown on an $Al_2O_3$ (11$\bar{2}$0) substrate. Sample I, deposited on an $Al_2O_3$ (10$\bar{1}$0) substrate shows two reflections at 30.00° and 62.32°, with $d$-spacing ration accurately 2, whereas sample F [also deposited on an $Al_2O_3$ (10$\bar{1}$0) substrate] exhibits only a single reflection at 62.25°. At the same time their Raman spectra are different, Fig. 6(d). Basically, the 104 and 208 reflections are quite strong in the prototype ilmenite and corundum structures. However, neither of the samples exhibited 30.12 reflection. Also the $a$ lattice parameter would be rather large, around 5.96 Å. The pattern of the sample F can be explained by assuming 300 orientation with a reasonable lattice parameter $a$ = 5.16 Å. This model, however, has a difficulty to explain the presence of the reflection at 30.0° and the absence of the peak at 101.9°, sample I. The results imply that the structure of the samples E, F and G and H and I is not the prototype corundum but is a derivative structure, as is also seen from Raman spectra. There are continuous changes in Raman spectra with decreasing Ti-content, as Figs. 5(a) and (b) show. The Raman peak approximately at 700 cm$^{-1}$, which can be assigned to a breathing oxygen octahedron $A_{1g}$ mode in $NiTiO_3$ (see, e.g., ref. 23), and the band at around 170 cm$^{-1}$ (both spectral areas shadowed) become gradually weaker with decreasing Ti-content and eventually these features are almost lacking from the spectrum collected on the sample E possessing no Ti. Interestingly, the only clear mode in samples E, F and G is centered at around 530 cm$^{-1}$, see Fig. 6(d). This peak is seen in all (Ni,Co)-rich samples, including samples C and D. We come back to this point below.

### 4.3. Structural model

XPS measurements showed that the valences were +2 for Ni and Co cations and +4 four Ti cations. This would violate the charge neutrality condition if only 2/3 of the octahedra were occupied. An evident way to preserve charge neutrality is to place "excess" $Co^{2+}$ and $Ni^{2+}$ cations into the vacant octahedra. Thus, every $Ti^{4+}$ cation replaced by a $Ni^{2+}/Co^{2+}$-cation results in an occupation of a vacant octahedron. This corresponds to a decrease in oxygen/cation ratio following a formula $A^{2+}_{1+2x}Ti^{4+}_{1-x}O_3$, where $A^{2+}$ is $Ni^{2+}/Co^{2+}$-cation. $x$ can also be negative, meaning that each "excess" Ti-cation is compensated for by forming a vacant octahedron, which is consistent with the features seen in Ti-rich films A and B. To gain further understanding the film compositions, including oxygen content, were determined through XPS measurements. As Fig. 6 shows, the measured O/(Ni+Co+Ti) ratios are close to the values expected for the vacancy filling model, especially in the Ti-poor region. In the Ti-rich area some deviation occurs, possibly due to the formation of $Ti^{3+}$. $Ti_2O_3$ has the corundum crystal structure in which all Ti-cations are at the $Ti^{3+}$ valence state. In terms of the proposed vacancy filling model samples E, F and G have a corundum-derived structure in which all octahedra are filled.

**Crystal symmetries.** Table 5 in Appendix 1 gives the structural parameters for a Model 3 in which all octahedra sites are filled by Ni and Co according to the composition ratios corresponding to samples E, F and G. The shift of the cations at the 12$c$ site was adjusted so that the measured intensity ratio, 0.16 between 006 and 0012 reflections was reproduced. The $z$-value, 0.349 is close to a typical value found in 3$d$-transition metal corundum compounds (compare to 0.352 given in ref. 2). A more general structural model, valid for compounds with Ti/(Ni+Co) ratio between 0 and 1 is obtained by gradually filling the octahedra sites in the ilmenite structure (space group $R\bar{3}$) by placing Ni, Co and Ti cations at 3$a$ and 3$b$ sites in their statistical ratios. In the Ti-rich range ($x$<0) the fraction of filled octahedra is smaller and possibly certain fraction of Ti are in a +3 valence state. Model 2 suits as an approximation for this range, as all layers contain all types of cations and also vacant octahedra and no cations are assumed at 3*a* or 3*b* sites.

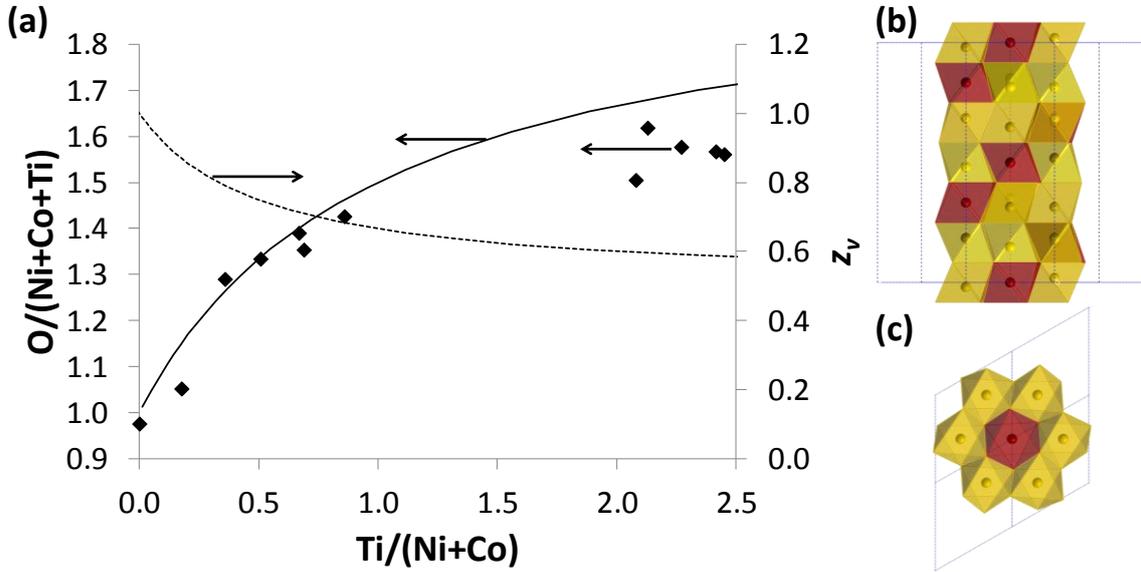

Fig. 6. (a) Measured oxygen per cation ratios as a function of Ti/(Ni+Co) ratio (filled diamonds). The solid line shows the O/(Ni+Co+Ti) ratio as a function of Ti/(Ni+Co) ratio corresponding to the formula $A_{1+2x}Ti_{1-x}O_3$. Dotted line gives the fraction of the filled octahedra $z_v$ as a function of Ti/(Ni+Co) ratio. (b) and (c) The corundum structure in which the octahedra indicated by red color are gradually filled with decreasing Ti/(Ni+Co) ratio. When the ratio is zero all octahedra are filled. Panel (b) gives an *bc*-plane projection and panel (c) shows an *ab*-plane projection.

In x-ray and neutron diffraction the gradual filling of the 3*a* and 3*b* sites in the ideal ilmenite and 6*b* in the ideal corundum structure does not change the reflection conditions, but changes the relative intensities. The symmetry change $R\bar{3} \to R\bar{3}c$ is obtained by displacing the two cations at 6*c* sites along the hexagonal *c* axis so that their fractional *z*-coordinates differ by ½ and displacing oxygen so that the fractional *y*-coordinate becomes zero. $R\bar{3}$ sites 3*a* and 3*b* together correspond to 6*b* site of the $R\bar{3}c$ symmetry. This symmetry change is probable since the compositions of the layers perpendicular to the *c* axis become identical when there is no titanium. The origin of the cation buckling in the ilmenite and corundum structures is related to the electrostatic forces displacing the cation asymmetrically along the *c* axis due to the vacant octahedra. Displacements are pictured in Figs. 1 and 6(b). Once all octahedra are filled the asymmetrical displacements expectably disappear, resulting in higher symmetry.

In order to understand the composition dependence of the reflection intensities at 30° and 62° and the gradual changes in Raman peak intensities a phase transformation sequence

$$R\bar{3} \to R\bar{3}c \to P6_3/mmc\ (2a, 2b, c) \to P6_3/mmc\ (a, b, c),$$

where $a = b \approx 2.980$ Å and $c \approx 4.870$ Å, is considered to illustrate symmetry changes following cation mixing and displacements and how they would be revealed by x-ray diffraction and Raman scattering measurements. The hexagonal lattice parameters of the ilmenite/corundum $(a', b', c')$ and the $P6_3/mmc\ (a, b, c)$ structures are related by equation $(a, b, c) = (a', b', c')M$, where
$M = \begin{bmatrix} -2/3 & 1/3 & 0 \\ -1/3 & -1/3 & 0 \\ 0 & 0 & 1/3 \end{bmatrix}$. Transformation $P6_3/mmc\ (2a, 2b, c) \to P6_3/mmc\ (a, b, c)$ corresponds to the halving of the *a* and *b* axes, without axes rotation. In the former phase there are two size octahedra, assigned to the $Ti^{4+}$ and $Ni^{2+}/Co^{2+}$ octahedra. We note that the symmetries on the left hand-side of the transformation sequence are probably lower as discussed in the case of sample J, though the present data are not sufficient for a precise space group assignment. The chain does not form a group-supergroup chain, though the symmetry increases from left to right. One could for instance speculate that there is a $P\bar{3}1c$ phase(s) after $R\bar{3}c$ in samples possessing Ti/(Ni+Co) ratios between samples D and H. We tried to find the simplest (the highest symmetry) structures explaining the observed features to give qualitative discussion.

In the $P6_3/mmc\ (2a,2b,c)$ phase the cations are placed at sites $2a(000)$ and $6g(½00)$ and oxygen are placed at sites $2c(⅓⅔¼)$ and $6h(x2x¼)$. When $x=⅚$ the $P6_3/mmc\ (2a,2b,c)$ phase reduces to $P6_3/mmc\ (a,b,c)$ phase in which cations are placed at the $2a$ and $2d(⅓⅔¾)$ sites. The Ti-content is decreasing from left to right. Though further studies, including electron diffraction studies, are required to understand the structure the following characteristics can be stated. The main structural changes corresponding to the symmetry changes are the mixing of cations, corresponding to the $R\bar{3} \to R\bar{3}c$ transformation. Once the fraction of filled octahedra increases the cation buckling (cation displacement towards vacant octahedron) decreases and eventually vanishes, corresponding to the $R\bar{3}c \to P6_3/mmc\ (2a,2b,c)$ transformation. The oxygen is still at position $6h$, which explains why a weak reflection is observed at 30°. Simulation shows that a weak 110 reflection is anticipated at 30.0°, a strong 220 reflection at 62.3° and vanishingly small 330 reflection at 101.7°. Fig. 7 shows the simulated pattern together with the experimental data, showing a good match between the measured and computed data. Also the crystal structure is shown by aligning the axes in such a way that the structural similarity with the CsCl-type structure is revealed. Since the lattice is hexagonal the cations are six fold coordinated, rather than eight fold coordinated as in the CsCl structure. Thus, the block formed by eight oxygen can be divided to two empty tetrahedra and one filled octahedron. One can speculate that the tetrahedral sites can be occupied in a mixed valence compounds, such as iron oxides commonly possessing iron at 2+ and 3+ valence. The size of the octahedron depends on the fractional coordinate parameter $x$ of the site $6h$ so that the red and gold colored octahedra in Fig. 7 are of different size. The network formed by oxygen is identical to the one found in the corundum and ilmenite structures, besides small displacements.

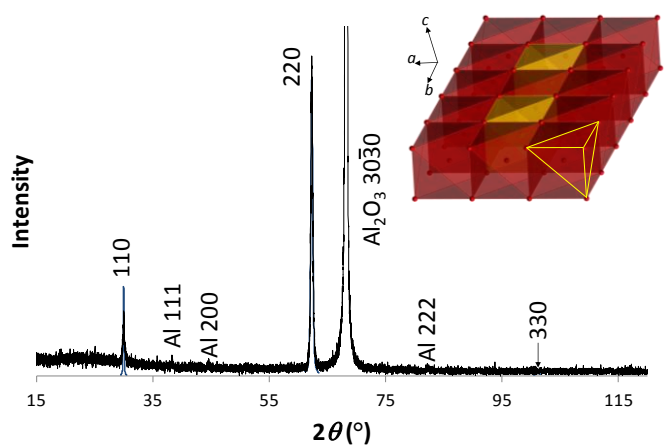

Fig. 7. Simulated x-ray diffraction intensity (blue line) based on the $P6_3/mmc\ (2a,2b,c)$ model. Experimental pattern was measured from the sample I. The intensities and peaks positions match with the experimental peak positions, reflections 110 and 220. Simulation predicts and experiment show vanishingly weak intensity for the 330 reflection. Also weak signal from the Al-sample-holder was detected. The lattice parameters used in the simulation were $a$ = 5.9560 Å and $c$ = 4.8703 Å. The parameter $x$ of the position $6h(x\ 2x\ ¼)$ was adjusted to 0.7675. The inset shows the structure. For clarity the layer above the shown layer, related by mirror plane perpendicular to the c-axis, is omitted. An empty tetrahedron is shown by yellow lines.

Symmetry analysis shows that the Raman active modes of this phase transform as $A_{1g}+3E_{2g}$. Once oxygen at 6$h$ position is shifted to the position (⅚⅔¼) a transformation $P6_3/mmc\ (2a,2b,c) \rightarrow P6_3/mmc\ (a,b,c)$ is completed. The $P6_3/mmc\ (a,b,c)$ phase has 110 reflection at 62.3°, corresponding to the pattern recorded from the sample F, Fig. 5(a). The $P6_3/mmc\ (a,b,c)$ phase has only one Raman active mode, which transforms as $E_{2g}$, consistently with the Raman spectra measured from samples E, F and G, Fig. 5(d). As a summary we make a structure assignment for all samples, Table 3.

Table 3. Summary of the crystal phases and estimates of the fraction of filled octahedra, $z_v$, estimated from Fig. 6(a). For the prototype ilmenite and corundum structures $z_v$ = 2/3.

| Sample | Phase | $z_v$ |
|---|---|---|
| A | Ilmenite[†] | 0.59 |
| B | Ilmenite[†] | 0.59 |
| C | Corundum[§] | 0.75 |
| D | Corundum[§] | 0.79 |
| E | $P6_3/mmc\ (a,b,c)$ | 1.00 |
| F | $P6_3/mmc\ (a,b,c)$ | 1.00 |
| G | $P6_3/mmc\ (a,b,c)$ | 1.00 |
| H | $P6_3/mmc\ (2a,2b,c)$ | 0.87 |
| I | $P6_3/mmc\ (2a,2b,c)$ | 0.87 |
| J | Ilmenite[‡] | 0.67 |

[†] Site occupancies as in Model 2, except that no cation site is fully occupied and the space group symmetry may be lower than $R\bar{3}$.
[§] Occupancies similar to Model 3, space group can differ from $R\bar{3}c$.
[‡] Model 1, except that the space group symmetry is lower than $R\bar{3}$. See text.

Since the cation/oxygen ratio significantly increases with increasing octahedra filling ratio both neutrons and x-rays suit for studying the structural evolution as illustrated by the simulations given in Appendix 3. Also electron diffraction studies should be used to address the symmetry changes.

### 4.4. Future work

Our preliminary squid measurements indicate that the films possess very clear remnant magnetization already at room temperature. We are currently studying magnetic and ferroelectric properties of the films. Magnetic properties depend on the configuration of cations and further on *x*. This implies that the models of magnetism based on the interactions derived for the ilmenite and corundum structures should be revised once *x*>0. For comparison, in *x*=0 compound only 1/3 of the octahedra are filled by Ni/Co, whereas every octahedra is filled by Ni/Co in *x*=1 compound. In other words the number of magnetic cations, in approximately the same volume, of the latter compound is 3 times the number of the former compound.

It is also worth checking if the sites vacant in the ideal corundum and ilmenite structures are partially filled in the hematite-ilmenite compounds. The magnetic properties of the hematite-ilmenite compounds are rather puzzling, exhibiting the so called self-reversed thermal remanence (21). Correspondingly, numerous experimental and theoretical studies and different interpretations can be found in the literature. Hematite-ilmenite compounds are a mixture of $Fe^{3+}$, $Fe^{2+}$ and $Ti^{4+}$ ions. Thus, once $Ti^{4+}$ cations are replaced by lower valence Fe-cations, a vacancy filling mechanism is one possibility.

**Conclusions**

Structural characteristics, especially the role of 3*d* orbital overlapping and the occupancy of octahedra, of nickel-cobalt-titanate (NCT) behind the functional properties were addressed. Single phase thin films were deposited by pulsed laser deposition. Thin films contrast bulk ilmenite oxides in two essential ways. First, films in which Ni, Co and Ti cations are mixed in the same layers were formed at relatively low temperatures. Second, both Ti-rich and Ni/Co-rich films were formed. In the former case more than one third of the octahedra are vacant, whereas in the latter case less than one third of the octahedra are vacant. In the ideal ilmenite and corundum structures one third of the octahedra are vacant. The composition of the films is $A_{1+2x}Ti_{1-x}O_3$, where $A$ is $Ni^{2+}$,$Co^{2+}$ and -0.25<$x$<1. The vacant octahedra sites can continuously be filled, which corresponds to a transformation of the ilmenite structure to the corundum-derived structure with all octahedra filled. The ilmenite-corundum phase transformation corresponds to the mixing of cations. A phase transformation sequence in which the loss of the cation displacements towards adjacent vacant octahedron and oxygen displacement in the hexagonal *ab*-plane gradually take place with decreasing Ti content was suggested. The sample with $x$ = 1 was assigned to space group symmetry $P6_3/mmc$ and has approximately 50 % larger cation density than the prototype ilmenite/corundum structures. The samples with $x \approx 0$ and approximately equal amounts of Ni and Co possess the ilmenite structure with symmetry lower than the prototype $R\overline{3}$. Magnetism and crystal distortion can be changed by adjusting $x$ and Ni/Co ratio: the crystal symmetry is decreasing with increasing Ti-content. Further efforts are dedicated to address the properties of NCT for multiferroic applications and to see if the same vacancy filling model occurs in related systems, such as the hematite-ilmenite compounds.

**Acknowledgments**

All experimental work was conducted at the Center for Nanophase Materials Sciences, which is a DOE Office of Science User Facility. We thank Dr. Jong Keum (Oak Ridge National Laboratory) for his help with XRD and XRR measurements. None of the authors have any competing interests in the manuscript.

# Appendix 1

Table 4. Structural parameters of the Models 1 and 2 for different films. Space group for both models is $R\bar{3}$. Axis setting is hexagonal. For Model 2, 003, 009, 0015, 003*n*, *n* is odd reflections vanish by symmetry. Occupancies were refined through XPS measurements.

|     |      |       |       |       | Model 1 Occupancy | | | Model 2 Occupancy | | |
| --- | ---- | ----- | ----- | ----- | ----- | ----- | ----- | ----- | ----- | ----- |
| Site | Atom | *x* | *y* | *z* | A | C | D | A | C | D |
| 6c | Ni1 | 0 | 0 | 0.642 | 0.352 | 0.362 | 0.348 | 0.168 | 0.220 | 0.232 |
|    | Co1 | 0 | 0 | 0.642 | 0.264 | 0.638 | 0.652 | 0.126 | 0.388 | 0.436 |
|    | Ti1 | 0 | 0 | 0.642 | 0.385 | 0 | 0 | 0.707 | 0.392 | 0.332 |
| 6c | Ni2 | 0 | 0 | 0.858 | 0 | 0.090 | 0.128 | 0.168 | 0.220 | 0.232 |
|    | Co2 | 0 | 0 | 0.858 | 0 | 0.158 | 0.241 | 0.126 | 0.388 | 0.436 |
|    | Ti2 | 0 | 0 | 0.858 | 1 | 0.752 | 0.631 | 0.707 | 0.392 | 0.332 |
| 18f | O | 0.305 | 0.015 | 0.250 | 1 | 1 | 1 | 1 | 1 | 1 |

Table 5. Structural parameters of the Model 3. Space group symmetry is $R\bar{3}c$. Axis setting is hexagonal. The *z* parameter of the atoms Ni1/Co1 was adjusted so that the intensity ratio between 006 and 0012 reflection corresponds to the experimental value. Occupancies were determined through XPS measurements.

| Site | Atom | *x* | *y* | *z* | Occupancy |
| --- | --- | --- | --- | --- | --- |
| 12c | Ni1 | 0 | 0 | 0.349 | 0.357 |
|     | Co1 | 0 | 0 | 0.349 | 0.574 |
| 6b  | Ni2 | 0 | 0 | 0 | 0.357 |
|     | Co2 | 0 | 0 | 0 | 0.574 |
| 18e | O | 0.305 | 0.0 | 0.250 | 1 |

**Appendix 2**

*FILM THICKNESS ESTIMATES*

## Method

The intensity of the scattered x-rays around each reciprocal-lattice point $s$ is proportional to $\frac{\sin^2(\pi N_a \boldsymbol{a}\cdot\boldsymbol{s})}{\sin^2(\pi \boldsymbol{a}\cdot\boldsymbol{s})} \frac{\sin^2(\pi N_b \boldsymbol{b}\cdot\boldsymbol{s})}{\sin^2(\pi \boldsymbol{b}\cdot\boldsymbol{s})} \frac{\sin^2(\pi N_c \boldsymbol{c}\cdot\boldsymbol{s})}{\sin^2(\pi \boldsymbol{c}\cdot\boldsymbol{s})}$. In the case of samples E and H the films are oriented so that the hexagonal $c$-axis is perpendicular to the substrate plane. The films are finite only in the $c$-axis direction, the thickness is $N_c c$. Fits to the XRD data collected on the films are given in Fig. 8 from which the thickness was estimated, Fig. 9.

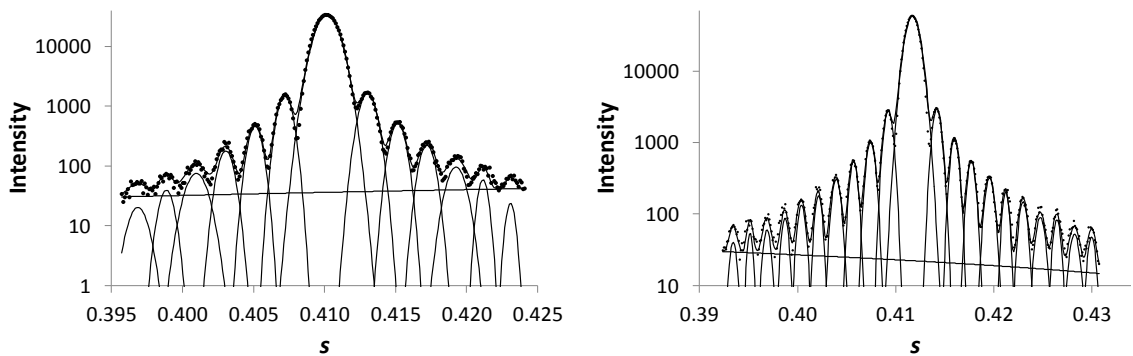

Fig. 8. Fit of the Gaussian functions to the x-ray diffraction intensity around the 006 reflection of the E (left) and H (right) samples.

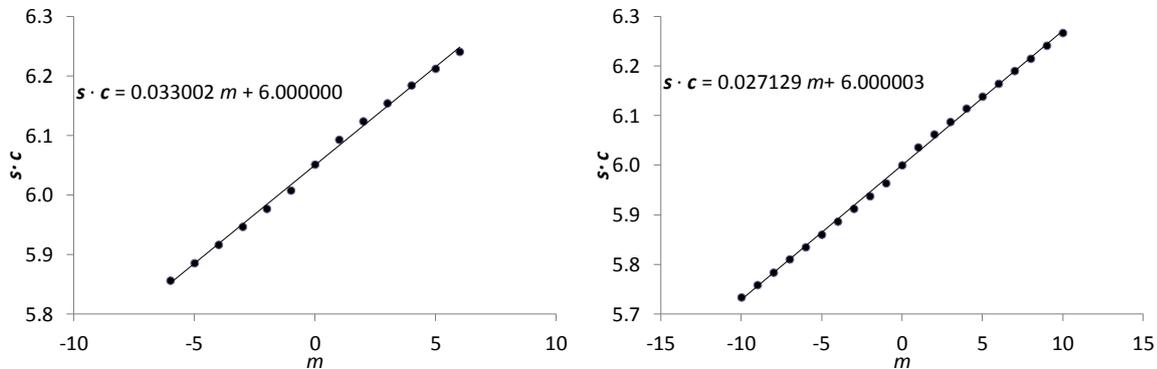

Fig. 9. The positions of subsidiary maxima measured from the samples E (left) and H (right) yield a film thickness estimates 44.3 and 53.7 nm, respectively.

**Appendix 3**

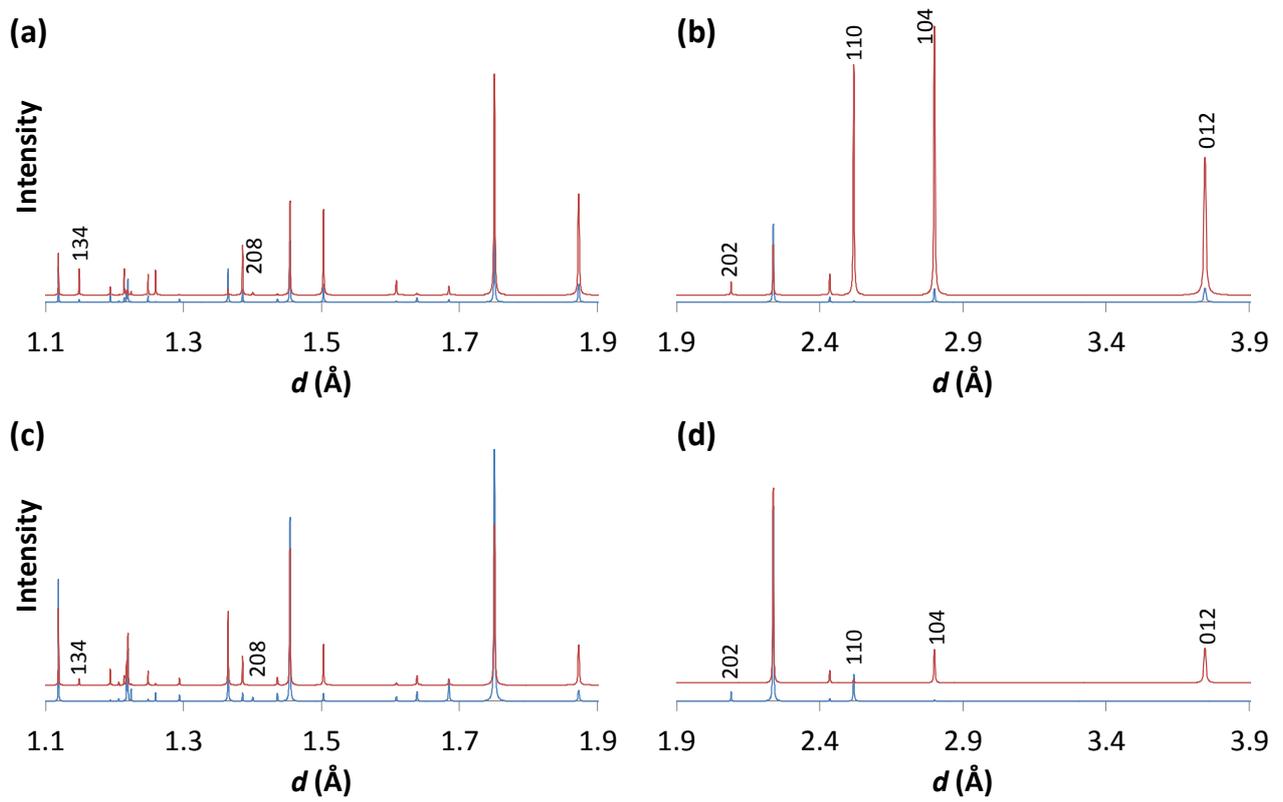

Fig. 10. Simulated x-ray diffraction, panels (a) and (b), and neutron powder diffraction, panels (c) and (d), patterns. Reflection conditions for the prototype corundum (red line) and all-octahedra-filled-corundum structure (blue line) are the same (both have the space group $R\bar{3}c$), but the relative intensities of the two structures are significantly different. The indexed reflections show cases in which one structure has nearly zero intensity, whereas the other has clear peaks. Also the complementary nature of x-rays and neutrons is seen.